\documentclass[10pt,aps,showpacs,twocolumn,pra,superscriptaddress]{revtex4-1}
\usepackage{amssymb}

\usepackage{amsmath}
\usepackage{amsmath,amssymb}
\usepackage{graphicx,color}
\usepackage{epstopdf}
\usepackage{epsfig}
\usepackage{mathrsfs}
\usepackage[breaklinks=true]{hyperref}

\input{tcilatex}

\begin{document}

\title{Interferometric Phase Estimation Though Quantum Filtering in Coherent States}
\author{John E.~Gough} \email{jug@aber.ac.uk}
   \affiliation{Institute for Mathematics and Physics, Aberystwyth University, SY23 3BZ, Wales, United Kingdom}

\begin{abstract}
We derive the form of the quantum filter equation describing the continuous
observation of the phase of a quantum system in an arm of an interferometer
via non-demolition measurements when the statistics of an input field used
for the indirect measurement are in a general coherent state. Both quadrature
homodyne detection and photon-counting dection schemes are covered, and we solve
the linearized filter for a specific application.
\end{abstract}

\maketitle

\section{Introduction}

There has been a steady interest in the problem of ``collapse of the wavefunction''
amongst quantum physicists, particularly in relation to foundational issues.
The dichotomy usually presented is between the unitary evolution under the
Schr\"{o}dinger equation and the discontinuous change when a measurement is made.
Clearly the collapse of the wavefunction is a form of conditioning the quantum 
state made by an instantaneous measurement. However, conditional probabilities
are well known classically and have no such interpretational issues. Furthermore,
the process of extraction of information from a classical system and the resulting
conditioning of the state is well studied from the point of view of stochastic
estimation. For continual measurements, there are standard results
on nonlinear filtering, see \cite{DavisMarcus,Kush79,Kush80,Zak69}. What is
not often appreciated in the theoretical physics community is that the analogue
problem was formulated by Belavkin \cite{Bel80,Bel92a} where a quantum theory of filtering based on
non-demolition measurements of an output field is established: see also \cite{BouGutMaa04}-\cite{GJNC_PRA12}. 
Specifically, we must measure a particular
feature of the field, for instance a field quadrature or the count of the
field quanta, and this determines a self-commuting, therefore \emph{%
essentially classical}, stochastic process. The resulting equations have
structural similarities with the classical analogues. They are also formally identical
with the equations arising in quantum trajectory theory \cite{Carmichael93} however the stochastic master
equations play different roles: in quantum filtering they describe the conditioned
evolution of the state while in quantum trajectories they are a means of simulating
a master equation.

\begin{figure}[tbph]
\centering
\includegraphics[width=0.40\textwidth]{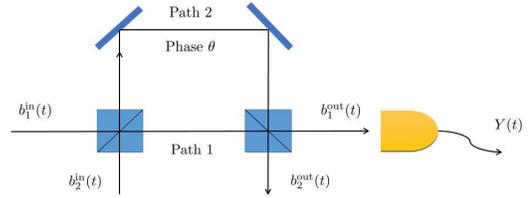}
\caption{The model is fully quantum: we have a Mach-Zehnder interferometer in
which there is a quantum mechanical phase associated with one of the arms (this may be due
to one or more of the mirrors being an opto-mechanical system); the input fields are
modelled a quantum input processes on the appropriate Boson Fock space (see Section \ref{sec:setup}).}
\label{fig:Interferometer}
\end{figure}

There has been recent interest amongst the physics community in quantum
filtering as an applied technique in quantum feedback and control \cite
{AASDM02}- \cite{WM93}. An
additional driver is the desire to go beyond the situation of a vacuum field
and derive the filter for other physically important states such as thermal,
squeezed, single photon states, etc. In a previous publication
\cite{G_scat_PRA15} we derived a quantum Markovian model for an opto-mechanical system consisting of a quantum
mechanical mirror interacting with quantum optical input fields via radiation pressure,
and in particular were able to construct the quantum filter for the position of the mirror
based on the continual monitoring of scattered photons. To obtain a non-trivial result,
we had to place the input fields in a coherent state of non-zero intensity and rely
on the filtering theory for coherent state inputs \cite{GK_COSA10}. In this note we wish to 
treat the problem of constructing the filter for non-demolition quadrature and photon-counting measurements 
of the output of a Mach-Zehnder interferometer with the purpose of estimating the phase difference
between the two arms of the interferometer: see Figure \ref{fig:Interferometer}. 

Here the phase is treated as a quantum mechanical
object, so the problem is genuinely one of estimating the quantum state of the interferometer phase
variable. As the interaction of the photons with the interferometer is purely scattering (so no
emission or absorption) we must take one of input fields to be in a coherent state with intensity
function $\beta $. The model may be thought of as the continuous variable analogue of the discrete
model examined recently by Harrell in \cite{Harrell}: indeed, it is reasonable to expect that 
the continuous time limit of this model leads to the results presented here by the time of
arguments presented in \cite{G_Sob_2004}.

The paper is organized as follows. In Section \ref{sec:setup} we describe the model of a Mach-Zehnder 
interferometer with appropriate continuous-variable quantum inputs. A fully quantum stochastic model of the
interferometer phase observable and the photon fields is presented in terms of quantum stochastic calculus
\cite{HP}-\cite{GC85}. In Section \ref{sec:filter} we describe the basic estimation problem and
state the main result which is the form of the filters in the language of stochastic estimation: these may then be 
rewritten in terms of stochastic master equations, and we give the equivalent form for homodyning.
In Section \ref{sec:derive_filters} we derive the filters using the characteristic function approach.
Finally in Section \ref{sec:Applications} we solve the filter in a linearized regime - equivalent to a 
quantum Kalman-Bucy filter and discuss the physical properties of the system including the ``collapse of
the wave-function''.

\section{The Experimental Set Up}
\label{sec:setup}
We consider an interferometer as shown in Figure \ref{fig:Interferometer}
where two continuous wave optical inputs $b_{1}^{\text{in}}$ and $b_{2}^{%
\text{in}}$ are mixed in a 50-50 beam splitter and then recombined in a
second 50-50 beam splitter. The second path in the interferometer has a
phase $\theta $ relative to the first path. We treat $\theta $ as a quantum
mechanical observable and we aim to estimate the corresponding state by
measuring one of the output fields. In this paper we will consider a
homodyne scheme where we measure the quadrature associated with the output $%
b_{1}^{\text{out}}$. The problem would of course be trivial if both inputs
where in the vacuum state, so we assume that one on the inputs, $b_{1}^{%
\text{in}}$ is in a coherent state while the other is in the vacuum.

The scattering matrix $S$ relating the inputs processes to outputs is given
by the product $S=TPT$ where $T=\frac{1}{\sqrt{2}}\left[ 
\begin{array}{cc}
1 & i \\ 
i & 1
\end{array}
\right] $ and $P=\left[ 
\begin{array}{cc}
1 & 0 \\ 
0 & -e^{i\theta }
\end{array}
\right] $ are the beam splitter matrix and interferometer path transfer
matrix respectively. That is 
\begin{eqnarray}
S &=&\left[ 
\begin{array}{cc}
S_{11} & S_{12} \\ 
S_{21} & S_{22}
\end{array}
\right]  \nonumber \\
&\equiv &\frac{1}{2}\left[ 
\begin{array}{cc}
1+e^{i\theta } & i\left( 1-e^{i\theta }\right) \\ 
i\left( 1-e^{i\theta }\right) & -\left( 1+e^{i\theta }\right)
\end{array}
\right] .  \label{eq:S_interferometer}
\end{eqnarray}
Note that the entries $S_{jk}$ depend on the observable $\theta $ and are
therefore operators on the associated Hilbert space $\mathfrak{h}$. We may
additionally have a Hamiltonian $H$ leading to a non-trivial evolution of
the observable $\theta $.

\subsection{Fully Quantum Model}

The inputs satisfy the singular commutation relations 
\begin{eqnarray*}
\left[ b_{j}^{\text{in}}\left( t\right) ,b_{k}^{\text{in*}}\left( s\right) %
\right] =\delta _{jk}\delta \left( t-s\right) ,
\end{eqnarray*}
and in the following we shall work with the processes 
\begin{eqnarray*}
B_{k}\left( t\right) &=&\int_{0}^tb_{k}^{\text{in}}\left( s\right) ds, \\
B_{k}\left( t\right) ^{\ast } &=&\int_{0}^tb_{k}^{\text{in*}}\left(
s\right) ds, \\
\Lambda _{jk}\left( t\right) &=&\int_{0}^tb_{j}^{\text{in*}}\left(
s\right) b_{k}^{\text{in}}\left( s\right) ds,
\end{eqnarray*}
which correspond to well defined operators on a Fock space $\mathfrak{F}$.
The Hudson-Parthasarathy theory of quantum stochastic calculus \cite{HP}
gives an analogue of the It\={o} calculus for integrals with respect to
these processes. We note the It\={o} table 
\begin{eqnarray}
dB_{j}dB_{k}^{\ast } &=&\delta _{jk}dt,  \nonumber \\
dB_{j}d\Lambda _{kl} &=&\delta _{jk}dB_{l},  \nonumber \\
d\Lambda _{jk}dB_{l}^{\ast } &=&\delta _{kl}dB_{j}^{\ast },  \nonumber \\
d\Lambda _{jk}d\Lambda _{lm} &=&\delta _{kl}d\Lambda _{jm},  \label{eq:QIT}
\end{eqnarray}
with other products of differentials vanishing.

The general class of unitaries processes on $\mathfrak{h}\otimes \mathfrak{F}
$ driven by these fundamental processes is given in \cite{HP} involves
coefficients $\left( S,L,H\right) $ corresponding to the scattering $S$, the
coupling $L$ and the Hamiltonian $H$ for the non-field component. In our
case their is no photo-emissive coupling of input fields with the
interferometer, so we set $L=0$. The most general form then reduces to 
\begin{eqnarray*}
dU_t=\left\{ \sum_{j,k}\left( S_{jk}-\delta _{jk}\right) \otimes d\Lambda
_{jk}\left( t\right) -iH\otimes dt\right\} U_t,
\end{eqnarray*}
with $S=\left[ S_{jk}\right] $ unitary, that is 
\begin{eqnarray}
\sum_{k}S_{ki}^{\ast }S_{kj}=\delta _{ij}I_{\mathfrak{h}}=%
\sum_{k}S_{ik}S_{jk}^{\ast }.  \label{eq:S_unitary}
\end{eqnarray}

\subsection{Internal Dynamics of the Interferometer}

For $X$ an arbitrary operator on the Hilbert space $\mathfrak{h}$ of the
interferometer, its Heisenberg evolution is given by 
\begin{eqnarray*}
j_t\left( X\right) =U_t^{\ast }\left( X\otimes I_{\mathfrak{F}}\right)
U_t
\end{eqnarray*}
and from the quantum It\={o} calculus we find Langevin equation 
\begin{eqnarray*}
dj_t\left( X\right)  &=&\sum_{i,j,k}j_t\left( S_{ki}^{\ast
}XS_{kj}-\delta _{ij}X\right) \otimes d\Lambda _{ij}\left( t\right)  \\
&&-ij_t\left( \left[ X,H\right] \right) \otimes dt.
\end{eqnarray*}
For the specific case of the scattering matrix (\ref{eq:S_interferometer})
we find 
\begin{eqnarray}
&&dj_t\left( X\right) =-ij_t\left( \left[ X,H\right] \right) \otimes dt 
\nonumber \\
&&+\frac{1}{2}j_t\left( e^{-i\theta }Xe^{i\theta }-X\right) \otimes \left(
d\Lambda _{11}-id\Lambda _{12}+id\Lambda _{21}+d\Lambda _{22}\right) .
\nonumber \\
\label{eq:Heisenberg}
\end{eqnarray}

\subsection{Input-Output Relations}

The output fields $B_{j}^{\text{out}}\left( t\right) $ are defined by 
\begin{eqnarray*}
B_{j}^{\text{out}}\left( t\right) \triangleq U_t^{\ast }\left( I_{%
\mathfrak{h}}\otimes B_{j}^{\text{in}}\left( t\right) \right) U_t
\end{eqnarray*}
and again using the quantum It\={o} calculus we find 
\begin{eqnarray*}
dB_{j}^{\text{out}}\left( t\right) =\sum_{k}j_t\left( S_{jk}\right)
\otimes dB_{k}^{\text{in}}\left( t\right) .
\end{eqnarray*}

\subsection{Homodyne Detection}

Our objective is to estimate the state of the interferometer at time $t$
based on the observations of the output quadrature $Y$ of the first output
up to time $t$. Here we have 
\begin{eqnarray*}
Y\left( t\right) &=&B_{1}^{\text{out}}\left( t\right) +B_{1}^{\text{out}%
}\left( t\right) ^{\ast }\nonumber \\
&=& U_t^{\ast }\left( I_{\mathfrak{h}}\otimes \left(
B_{1}^{\text{in}}\left( t\right) +B_{1}^{\text{in}}\left( t\right) ^{\ast
}\right) \right) U_t.
\end{eqnarray*}
We note that 
\begin{eqnarray}
dY\left( t\right) =\sum_{k}j_t\left( S_{1k}\right) \otimes dB_{k}^{\text{in%
}}\left( t\right) +H.c.  \label{eq:dY_quadrature}
\end{eqnarray}
The process $Y=Y^{\ast }$ is self-nondemolition by which we mean that $\left[
Y\left( t\right) ,Y\left( s\right) \right] =0$ for all times $t,s$. It
furthermore satisfies the nondemolition property that $\left[ j_t\left(
X\right) ,Y\left( s\right) \right] =0$ for all $t\geq s$, so that we may
estimate present or future values of the observable $X$ in the Heisenberg
picture based on the observations up to and including present time. We note
that 
\begin{eqnarray*}
\left( dY\right) ^{2}=dt,
\end{eqnarray*}
which follows from the quantum It\={o} table (\ref{eq:QIT}) and the
unitarity condition (\ref{eq:S_unitary}).

Clearly the process $Y$ contains information about the scattering
coefficients, however, it would possess the statistics of a standard Wiener
process if we took the input fields to be in the vacuum state. It is for
this reason we take the first input field to be in a coherent state
corresponding to an intensity $\beta =\beta \left( t\right) $. The joint
state is denoted as $\mathbb{E}_{\beta }$ and is the product state of the
initial state of the interferometer (which may be a guess!) and the Gaussian
state of the fields corresponding to input 1 in the coherent state with
intensity $\beta $ and input 2 in the vacuum. Specifically, the Weyl
operators have expectation 
\begin{multline*}
\mathbb{E}_{\beta }\left[ e^{\sum_{k}\int f_{k}\left( t\right) dB_{k}^{\text{%
in}}\left( t\right) ^{\ast }-H.c.}\right] =e^{-\frac{1}{2}\sum_{k}\int
|f_{k}\left( t\right) |^{2}dt} \\
\times e^{\int f_{1}\left( t\right) \beta \left( t\right) ^{\ast }dt-\int
f_{1}\left( t\right) ^{\ast }\beta \left( t\right) dt},
\end{multline*}
and so 
\begin{eqnarray}
\mathbb{E}_{\beta }\left[ dY\left( t\right) \right] =\mathbb{E}_{\beta }%
\left[ j_t\left( S_{11}\right) \right] \beta \left( t\right) dt+\mathbb{E}%
_{\beta }\left[ j_t\left( S_{11}^{\ast }\right) \right] \beta \left(
t\right) ^{\ast }dt \nonumber\\
 \label{eq:E[dY]}
\end{eqnarray}
which is non-zero for $\beta \left( t\right) \neq 0$.

\subsection{Photon Counting Detection}

Alternatively we could count the number of photons at the first output
channel. This time the measured process $Y$ is 
\begin{eqnarray*}
Y\left( t\right) =U_t^{\ast }\left( I_{\mathfrak{h}}\otimes \Lambda _{11}^{%
\text{out}}\left( t\right) \right) U_t
\end{eqnarray*}
and from the It\={o} calculus we obtain 
\begin{eqnarray}
dY\left( t\right)  &=&\sum_{j,k}j_t\left( S_{1j}^{\ast }S_{1k}\right)
\otimes d\Lambda _{jk}^{\text{in}}\left( t\right)   \nonumber \\
&=&j_t\left( \frac{1+\cos \theta }{2}\right) \otimes d\Lambda _{11}^{\text{%
in}}\left( t\right)   \nonumber \\
&&+j_t\left( \frac{\sin \theta }{2}\right) \otimes \left[ d\Lambda _{12}^{%
\text{in}}\left( t\right) +d\Lambda _{21}^{\text{in}}\left( t\right) \right] 
\nonumber \\
&&+j_t\left( \frac{1-\cos \theta }{2}\right) \otimes d\Lambda _{22}^{\text{%
in}}\left( t\right) .  \label{eq:dY_photon_count}
\end{eqnarray}

Here we note that 
\begin{eqnarray}
\left( dY\right) ^{2}=j_t\left( \frac{1+\cos ^{2}\theta }{2}\right)
\otimes d\Lambda _{11}^{\text{in}}\left( t\right) +\cdots
\label{eq:dY^2_photon_count}
\end{eqnarray}
where the omitted terms are proportional to the increments $d\Lambda _{12}^{%
\text{in}}\left( t\right) ,d\Lambda _{21}^{\text{in}}\left( t\right) $ and $%
d\Lambda _{22}^{\text{in}}\left( t\right) $ which average to zero of the
state $\mathbb{E}_{\beta }$.

\section{Quantum Filtering}
\label{sec:filter}
Our goal is to derive the optimal estimate $\pi _t\left( X\right) $ for an
observable $j_t\left( X\right) $ for the state $\mathbb{E}_{\beta }$ given
the observations of $Y$ up to time $t$. To this end we set $\mathfrak{Y}%
_{t]} $ to be the (von Neumann) algebra generated by the family $\left\{
Y\left( s\right) :0\geq s\leq t\right\} $. As $Y$ is self-non-demolition, we
have that $\mathfrak{Y}_{t]}$ is a commutative algebra - so the recorded
measurement can be treated as an essentially classical stochastic process as
it should be. Every observable that commutes with $\mathfrak{Y}_{t]}$ will
possess a well-defined joint (classical) statistical distribution with the
measurements up to time $t$ and by the non-demolition property this includes 
$j_t\left( X\right) $. We therefore set 
\begin{eqnarray*}
\pi _t\left( X\right) =\mathbb{E}_{\beta }\left[ j_t\left( X\right) |%
\mathfrak{Y}_{t]}\right]
\end{eqnarray*}
which is the conditional expectation of $j_t\left( X\right) $ onto $%
\mathfrak{Y}_{t]}$. The right hand side always exists since the algebra
generated by $\mathfrak{Y}_{t]}$ and the additional element $j_t\left(
X\right) $ is commutative and so we exploit the fact that in classical
probability theory conditional expectations always exist. This classical
expectation is then understood as being a function of the commuting set $%
\left\{ Y\left( s\right) :0\geq s\leq t\right\} $. It should be remembered
that $\pi _t\left( X\right) $ can be defined in this way for arbitrary
observable $X$ of the interferometer system, and these $X$'s generally do
not commute, so the construction is genuinely quantum in that regard. Note
that while $\pi _t\left( X_{1}X_{2}\right) $ is generally different from $%
\pi _t\left( X_{2}X_{1}\right) $, we will however have $\left[ \pi
_t\left( X_{1}\right) ,\pi _t\left( X_{2}\right) \right] =0$ as we have
conditioned onto the commutative algebra of operators $\mathfrak{Y}_{t]}$.
Finally we mention that this estimate is optimal in the least squares sense.
That is,for $X$ self-adjoint, we have 
\begin{eqnarray*}
\mathbb{E}_{\beta }\bigg[\left( j_t\left( X\right) -\pi _t\left(
X\right) \right) ^{2}\bigg]\leq \mathbb{E}_{\beta }\bigg[\left( j_t\left(
X\right) -\hat{X}_t\right) ^{2}\bigg]
\end{eqnarray*}
for all $\hat{X}_t\in \mathfrak{Y}_{t]}$. In particular we have the
``orthogonality'' property

\begin{eqnarray*}
\mathbb{E}_{\beta }\bigg[\left( j_t\left( X\right) -\pi _t\left(
X\right) \right) C\left( t\right) \bigg]=0
\end{eqnarray*}
for every $C\left( t\right) \in \mathfrak{Y}_{t]}$.

\subsection{The Filtering Equation}
We will now state the main result. In both cases the filtering equation takes the form
\begin{eqnarray}
d\pi _t\left( X\right)  &=&\frac{1}{2}\pi _t\left( e^{-i\theta
}Xe^{i\theta }-X\right) |\beta \left( t\right) |^{2}-i\pi _t(\left[ X,H%
\right] )dt  \nonumber \\
&&+\mathcal{H}_t\left( X\right) \,dI\left( t\right) ,
\label{eq:interf_filter}
\end{eqnarray}
The terms $\mathcal{H}_t(X)$ and the process $I(t)$ are specific to the physical 
mode of detection and we give these explicitly for the scheme of homodyne (quadrature)
measurement and the photon counting scheme below.

\subsubsection{Quadrature Measurement}
In this case we measure the quadrature of output field 1. here we find that
\begin{eqnarray}  
\mathcal{H}_t\left( X\right) &=&\frac{1}{2}\left[ \pi _t\left(
Xe^{i\theta }\right) -\pi _t\left( X\right) \pi _t\left( e^{i\theta
}\right) \right] \beta \left( t\right)  \nonumber \\
&+& \frac{1}{2}\left[ \pi _t\left( e^{-i\theta }X\right) -\pi _t\left(
e^{-i\theta }\right) \pi _t\left( X\right) \right] \beta \left( t\right)
^{\ast }.  \nonumber \\
\label{eq:H_final_quad}
\end{eqnarray}
The innovations process $I$ is defined by 
\begin{eqnarray}
dI\left( t\right) =dY\left( t\right) -\left[ \pi _t\left( S_{11}\right)
\beta \left( t\right) +\pi _t\left( S_{11}^{\ast }\right) \beta \left(
t\right) ^{\ast }\right] dt, \nonumber \\ \label{eq:Innovations}
\end{eqnarray}
and $I\left( 0\right) =0$. 
Statistically it has the distribution of a standard Wiener process.

We see that $dI\left( t\right) $ is the difference between
the actual observed increment $dY\left( t\right) $ and the expected
increment $\left[ \pi _t\left( S_{11}\right) \beta \left( t\right) +\pi
_t\left( S_{11}^{\ast }\right) \beta \left( t\right) ^{\ast }\right] dt$
based on the filter. In stochastic estimation problems $I(t)$ is referred to
as the innovations process.

\subsubsection{Photon Counting Measurement}
If instead we count the photons coming out of output 1, we obtain
\begin{eqnarray}
 \mathcal{H}_t (X) &=& \frac{ 1  }{1+\pi _t\left( \cos ^{2}\theta \right) }
\nonumber
\\
&&  \times \Big\{ \frac{1}{2}\pi _t\left( e^{-i\theta }Xe^{i\theta
}+e^{-i\theta }X+Xe^{i\theta }-X\right) \nonumber \\
&& \quad \quad \qquad -  \pi _t\left( X\right) \pi
_t\left( \cos \theta \right)   \Big\} .
\label{eq:H_photon_count}
\end{eqnarray}
The innovations process is this time given by
\begin{eqnarray}
dI\left( t\right) =dY\left( t\right) -\pi _t\left( \frac{1+\cos \theta }{2}%
\right) |\beta \left( t\right) |^{2}dt,  \label{eq:Innovations_photon_count}
\end{eqnarray}
and $I\left( 0\right) =0$. This time the innovations have the statistical
distribution of a compensated Poisson process. Once again, $dI(t)$ is the
difference between the observed measurement increment $dY(t)$ and the
expected increment $\mathbb{E}_\beta [ dY(t) ]$.

\subsubsection{Equivalent Stochastic Master Equation}

We may alternatively use the dual form for the filter where we express
everything in terms of the conditional state $\varrho _t$ of the
interferometer system based on the measurements so that 
\begin{eqnarray*}
\pi _t\left( X\right) =\text{tr}\left\{ \varrho _tX\right\} .
\end{eqnarray*}
In the quadrature case, the filter equation thereby translates into the equivalent stochastic master
equation (SME) for $\varrho _t$%
\begin{eqnarray}
d\varrho _t &=&\frac{1}{2}\left( e^{i\theta }\varrho _te^{-i\theta
}-\varrho _t\right) |\beta \left( t\right) |^{2}dt+i\left[ \varrho _t,H%
\right] dt  \nonumber \\
&&+\frac{1}{2}\left( e^{i\theta }\varrho _t-\zeta _t\varrho _t\right)
\beta \left( t\right) dI\left( t\right)  \nonumber \\
&&+\frac{1}{2}\left( \varrho _te^{-i\theta }-\zeta _t^{\ast }\varrho
_t\right) \beta \left( t\right) ^{\ast }dI\left( t\right)  \label{eq:SME}
\end{eqnarray}
where 
\begin{eqnarray}
\zeta _t=\text{tr}\left\{ \varrho _te^{i\theta }\right\} .
\label{eq:zeta}
\end{eqnarray}

The presence of the term (\ref{eq:zeta}) in the SME (\ref{eq:SME}) means
that the equation is nonlinear. This filter is diffusive since the
innovations are a standard Wiener process. An SME formulation may likewise be
given for the photon counting: this again will be nonlinear, but this time 
will be driven by a jump process corresponding to the observation of a
photon arrival at the detector.

If we choose to ignore the measurement
record - a nonselective measurement - then we obtain the following master
equation for $\bar{\rho}_t=\mathbb{E}_{\beta }\left[ \varrho _t\right] $%
: 
\begin{eqnarray}
\frac{d\bar{\rho}_t}{dt}=\frac{1}{2}\left( e^{i\theta }\bar{\rho}%
_te^{-i\theta }-\bar{\rho}_t\right) |\beta \left( t\right) |^{2}+i\left[ 
\bar{\rho}_t,H\right] .  \label{eq:ME}
\end{eqnarray}
In the language of quantum trajectories, the SME (\ref{eq:SME}) is an
``unravelling'' of the master equation (\ref{eq:ME}). The same master
equation is unravelled by the photon counting SME.

\section{Derivation of the filters}
\label{sec:derive_filters}
In this section we derive the form of the filters given in Section 
\ref{sec:filter}. We will use a technique known as the \textit{characteristic function approach}.
This is a direct method for calculating the filtered estimate $\pi _t\left(
X\right) $ is based on introducing a process $C\left( t\right) $ satisfying
the QSDE 
\begin{eqnarray}
dC\left( t\right) =f\left( t\right) C\left( t\right) dY\left( t\right) ,
\end{eqnarray}
with initial condition $C\left( 0\right) =I$. Here we assume that $f$ is
integrable, but otherwise arbitrary. The technique is to make an ansatz of
the form 
\begin{eqnarray}
d\pi _t\left( X\right) =\mathcal{F}_t\left( X\right) dt+\mathcal{H}%
_t\left( X\right) dY\left( t\right)  \label{filter ansatz}
\end{eqnarray}
where we assume that the processes $\mathcal{F}_t\left( X\right) $ and $%
\mathcal{H}_t\left( X\right) $ are adapted and lie in $\mathfrak{Y}_{t]}$.
These coefficients may be deduced from the identity 
\begin{eqnarray*}
\mathbb{E}\left[ \left( \pi _t\left( X\right) -j_t\left( X\right)
\right) C\left( t\right) \right] =0
\end{eqnarray*}
which is valid since $C\left( t\right) \in \mathfrak{Y}_{t]}$. We note that
the It\={o} product rule implies $I+II+III=0$ where

\begin{eqnarray*}
I &=&\mathbb{E}\left[ \left( d\pi _t\left( X\right) -dj_t\left( X\right)
\right) C\left( t\right) \right] , \\
II &=&\mathbb{E}\left[ \left( \pi _t\left( X\right) -j_t\left( X\right)
\right) dC\left( t\right) \right] , \\
III &=&\mathbb{E}\left[ \left( d\pi _t\left( X\right) -dj_t\left(
X\right) \right) dC\left( t\right) \right] .
\end{eqnarray*}

\subsection{The Quadrature Filter}

We now compute the filter when $Y$ is the measured quadrature of the first
output channel.

\subsubsection{Term I}

Here we have (omitting $t$-dependence for ease of notation)

\begin{multline*}
I=\mathbb{E}_{\beta }\left[ \mathcal{F}\left( X\right) C+\mathcal{H}\left(
X\right) (j(S_{11})\beta +j(S_{11}^{\ast })\beta ^{\ast })C\right] dt \\
-\frac{1}{2}\mathbb{E}_{\beta }\left[ j\left( e^{-i\theta }Xe^{i\theta
}-X\right) C\right] |\beta |^{2}dt+i\mathbb{E}_{\beta }\left[ \left[ X,H%
\right] C\right] dt
\end{multline*}
where we use the fact that 
\begin{eqnarray*}
\mathbb{E}_{\beta }\left[ d\Lambda _{11}\left( t\right) \right] =|\beta
\left( t\right) |^{2}dt
\end{eqnarray*}
while $\mathbb{E}_{\beta }\left[ d\Lambda _{jk}\left( t\right) \right] =0$
otherwise.

\subsubsection{Term II}

From (\ref{eq:E[dY]}) we obtain

\begin{eqnarray*}
II=f\,\mathbb{E}_{\beta }\left[ \left( \pi \left( X\right) -j(X)\right)
C\left( t\right) j(S_{11})\beta +j(S_{11}^{\ast })\beta ^{\ast }\right] dt.
\end{eqnarray*}

\subsubsection{Term III}

We have

\begin{multline*}
III=f\,\mathbb{E}_{\beta }\left[ \mathcal{H}\left( X\right) C\right] dt \\
-f\,\frac{1}{2}\mathbb{E}_{\beta }\left[ j\left( e^{-i\theta }Xe^{i\theta
}-X\right) \left( j\left( S_{11}^{\ast }\right) -ij\left( S_{12}^{\ast
}\right) \right) \beta ^{\ast }C\right] dt
\end{multline*}
where we use the fact that $\left( dY\right) ^{2}=dt$ and from (\ref
{eq:Heisenberg}) and (\ref{eq:dY_quadrature}) 
\begin{eqnarray*}
dj\left( X\right) dY &=&\frac{1}{2}j_t\left( e^{-i\theta }Xe^{i\theta
}-X\right)  \\
&&\times \left( j\left( S_{11}^{\ast }\right) -ij\left( S_{12}^{\ast
}\right) \right) dB_{1}^{\ast }+\cdots 
\end{eqnarray*}
where the omitted terms average to zero. Note that we used the identities $%
d\Lambda _{11}dB_{1}^{\ast }=d\Lambda _{12}dB_{2}^{\ast }=dB_{1}^{\ast }$.

\subsubsection{Computing the Filter}
Now from the identity $I+II+III=0$ we may extract separately the
coefficients of $f\left( t\right) C\left( t\right) $ and $C\left( t\right) $
as $f\left( t\right) $ was arbitrary to deduce 
\begin{multline*}
\pi \left( \left( \pi \left( X\right) -j(X)\right) \left( j(S_{11})\beta
+j(S_{11}^{\ast })\beta ^{\ast }\right) \right) + \pi (\mathcal{H}\left( X\right)) \\
-\pi \left( \frac{1}{2}j\left( e^{-i\theta
}Xe^{i\theta }-X\right) \left( j\left( S_{11}^{\ast }\right) -ij\left(
S_{12}^{\ast }\right) \right) \beta ^{\ast }\right) =0,
\end{multline*}
and
\begin{multline*}
0=\pi \left( \mathcal{F}\left( X\right) +\mathcal{H}\left( X\right)
(j(S_{11})\beta +j(S_{11}^{\ast })\beta ^{\ast }\right) \\
-\frac{1}{2}\pi \left( e^{-i\theta }Xe^{i\theta }-X\right) |\beta |^{2}+i\pi
(\left[ X,H\right] ).
\end{multline*}
Using the projective property of the conditional expectation $\left( \pi
_t\circ \pi _t=\pi _t\right) $ and the assumption that $\mathcal{F}%
_t\left( X\right) $ and $\mathcal{H}_t\left( X\right) $ already lie in $%
\mathfrak{Y}_{t]}$, we find after a little algebra that 
\begin{eqnarray}
\mathcal{H}_t &=&\left[ \pi _t\left( XS_{11}\right) -\pi _t\left(
X\right) \pi _t\left( S_{11}\right) \right] \beta \left( t\right)  \nonumber
\\
&&+[\frac{1}{2}\pi _t\left( \left( e^{-i\theta }Xe^{i\theta }+X\right)
S_{11}^{\ast }\right) -\pi _t\left( X\right) \pi _t\left( S_{11}^{\ast
}\right)  \nonumber \\
&&-\frac{i}{2}\pi _t\left( \left( e^{-i\theta }Xe^{i\theta }-X\right)
S_{12}^{\ast }\right) ]\beta \left( t\right) ^{\ast },  \label{eq:H_filter}
\\
\mathcal{F}_t &=&\frac{1}{2}\pi _t\left( e^{-i\theta }Xe^{i\theta
}-X\right) |\beta |^{2}-i\pi _t(\left[ X,H\right] )  \nonumber \\
&&-\mathcal{H}_t\left( X\right) \left[ \pi _t\left( S_{11}\right) \beta
\left( t\right) +\pi _t\left( S_{11}^{\ast }\right) \beta \left( t\right)
^{\ast }\right] .  \label{eq:F_filter}
\end{eqnarray}

Inserting the expressions $S_{11}\equiv \frac{1}{2}\left( 1+e^{i\theta
}\right) $ and $S_{12}=\frac{i}{2}\left( 1-e^{i\theta }\right) $ into (\ref
{eq:H_filter}) leads to the more symmetric form (\ref{eq:H_final_quad}).

Substituting the identity (\ref{eq:F_filter}) into the equation (\ref{filter ansatz}),
$d\pi _t\left( X\right) =\mathcal{F}_t\left( X\right) dt
+\mathcal{H}_t\left( X\right) dY\left( t\right)$, we arrive explicitly at (\ref{eq:interf_filter})
where $\mathcal{H}_t\left( X\right) $ is given by (\ref{eq:H_filter}) and
the process $I(t)$ is defined as in (\ref{eq:Innovations}).

Comparing with (\ref{eq:E[dY]}), we see that the
process $I$ is mean-zero for the state $\mathbb{E}_{\beta }$ and satisfies
the property $\left( dI\right) ^{2}=dt$. By L\'{e}vy's characterization theorem, it is a
Wiener process: see for instance \cite{Rogers_Williams} Theorem 33.1.

\subsection{The Photon Count Filter}

We now compute the filter when $Y$ is the measured photon count of the first
output channel. Again we omit $t$-dependence for ease of notation.

\subsubsection{Term I}

This time we have using (\ref{eq:dY_photon_count}) and (\ref{eq:E[dY]})

\begin{multline*}
I=\mathbb{E}_{\beta }\left[ \mathcal{F}\left( X\right) C+\mathcal{H}\left(
X\right) j(\frac{1+\cos \theta }{2})|\beta |^{2}C\right] dt \\
-\frac{1}{2}\mathbb{E}_{\beta }\left[ j\left( e^{-i\theta }Xe^{i\theta
}-X\right) C\right] |\beta |^{2}dt+i\mathbb{E}_{\beta }\left[ \left[ X,H%
\right] C\right] dt.
\end{multline*}
.

\subsubsection{Term II}

From (\ref{eq:dY_photon_count}) and (\ref{eq:E[dY]}) we obtain

\begin{eqnarray*}
II=f\,\mathbb{E}_{\beta }\left[ \left( \pi \left( X\right) -j(X)\right)
C\left( t\right) j(\frac{1+\cos \theta }{2})\right] |\beta |^{2}dt.
\end{eqnarray*}

\subsubsection{Term III}

We have

\begin{multline*}
III=f\,\mathbb{E}_{\beta }\left[ \mathcal{H}\left( X\right) j\left( \frac{%
1+\cos \theta }{2}\right) C\right] \left| \beta \right| ^{2}dt \\
-f\,\frac{1}{4}\mathbb{E}_{\beta }\left[ j\left( e^{-i\theta }Xe^{i\theta
}+e^{-i\theta }X-Xe^{i\theta }-X\right) C\right] \left| \beta \right| ^{2}dt
\end{multline*}
where we now use (\ref{eq:dY^2_photon_count}), and the fact that from (\ref
{eq:Heisenberg}) and (\ref{eq:dY_photon_count})
\begin{eqnarray*}
dj\left( X\right) dY &=&\frac{1}{2}j_t\left( e^{-i\theta }Xe^{i\theta
}-X\right)  \\
&&\times \left( j\left( S_{11}^{\ast }\right) -ij\left( S_{12}^{\ast
}\right) \right) dB_{1}^{\ast }+\cdots 
\end{eqnarray*}
where the omitted terms average to zero.

\subsubsection{Computing the Filter}

Collecting the coefficients of $f\left( t\right) C\left( t\right) $ and $%
C\left( t\right) $ as $f\left( t\right) $ from the identity $I+II+III=0$, we
now obtain the expression $\mathcal{H}_t\left( X\right)$ (\ref{eq:H_photon_count}) and
\begin{eqnarray}
\mathcal{F}_t (X) &=&\frac{1}{2}\pi _t\left( e^{-i\theta }Xe^{i\theta
}-X\right) |\beta |^{2}-i\pi _t(\left[ X,H\right] )  \nonumber \\
&&-\mathcal{H}_t\left( X\right) \pi _t\left( \frac{1+\cos \theta }{2}%
\right) |\beta \left( t\right) |^{2}.  \label{eq:F_filter_photon}
\end{eqnarray}

Substituting this into the equation (\ref{filter ansatz}) gives the stated result.

\section{Collapse of the Wavefunction}
\label{sec:Applications}
We shall follow \cite{Harrell} and set 
\begin{eqnarray*}
\theta =2kq+\pi \left( 2n+\frac{1}{2}\right)
\end{eqnarray*}
and for $k$ small we make the linearization 
\begin{eqnarray*}
e^{i\theta }\approx i-2kq.
\end{eqnarray*}
Under this approximation the stochastic master equation becomes linear and
we have 
\begin{eqnarray*}
\mathcal{H}_t\left( X\right) &=&-k\left[ \pi _t\left( Xq\right) -\pi
_t\left( X\right) \pi _t\left( q\right) \right] \beta \left( t\right) \\
&&-k\left[ \pi _t\left( qX\right) -\pi _t\left( X\right) \pi _t\left(
q\right) \right] \beta \left( t\right) ^{\ast }.
\end{eqnarray*}
If we assume that interferometer is internally static (that is, we take the
Hamiltonian $H=0$) then for functions of the observable $q$ we get 
\begin{eqnarray*}
d\pi _t\left( f\left( q\right) \right) &=& -k\left[ \pi _t\left( f\left(
q\right) q\right) -\pi _t\left( f\left( q\right) \right) \pi _t\left(
q\right) \right]  \nonumber \\
&&\times \left( \beta \left( t\right) +\beta \left( t\right) ^{\ast }\right)
dI\left( t\right) .
\end{eqnarray*}
So we find $d\pi _t\left( q\right) =-k\mathscr{V}_t\left( \beta \left( t\right) +\beta \left(
t\right) ^{\ast }\right) dI\left( t\right)$, where 
\begin{eqnarray}
\mathscr{V}_t\triangleq \pi _t\left( q^{2}\right) -\pi _t\left( q\right) ^{2}.
\label{eq:var_q}
\end{eqnarray}
We note that $\mathscr{V}_t$ is the conditional variance of the observable $q$.

The filter equation for the observable $q$ is of Kalman-Bucy form. In
such cases, if the initial state implies a Gaussian distribution for $q$,
then classically one expects the Gaussianity to be maintained and that the
variance $\mathscr{V}_t$ is deterministic. One will then have the property that all
moments may be expressed in terms of first and second moments, and in
particular third order moments of jointly Gaussian observables $X,Y,Z$ may
be rewritten as 
\begin{multline}
\pi _{t}(XYZ)\equiv   \nonumber \\
\pi _{t}(X)\pi _{t}(YZ)+\pi _{t}(Y)\pi _{t}(XZ)+\pi _{t}(Z)\pi _{t}(XY) 
\nonumber \\
-2\pi _{t}(X)\pi _{t}(Y)\pi _{t}(Z).  
\end{multline}
\begin{eqnarray}
\label{eq:3rd moments}
\end{eqnarray}
We will now show that this applies in the present situation.

We see that 
\begin{eqnarray*}
d\pi _t\left( q^{2}\right)  &=&-k\left[ \pi _t\left( q^{3}\right) -\pi
_t\left( q^{2}\right) \pi _t\left( q\right) \right]  \\
&&\left( \beta \left( t\right) +\beta \left( t\right) ^{\ast }\right)
dI\left( t\right) 
\end{eqnarray*}
however if the conditional distribution is Gaussian then we may use (\ref
{eq:3rd moments}) to write the third moment $\pi _t\left( q^{3}\right) $
as 
\begin{eqnarray*}
\pi _t\left( q^{3}\right) \equiv 3\pi _t\left( q\right) \pi _t\left(
q^{2}\right) -2\pi _t\left( q\right) ^{3}
\end{eqnarray*}
so that $d\pi _t\left( q^{2}\right) \equiv -2k\mathscr{V}_t\pi _t\left( q\right)
\left( \beta \left( t\right) +\beta \left( t\right) ^{\ast }\right) dI\left(
t\right) $. Applying the It\={o} calculus, recall $\left( dY\right) ^{2}=dt$%
, we have 
\begin{eqnarray*}
d\mathscr{V}_t &=&d\pi _t\left( q^{2}\right) +2\pi _t\left( q\right) d\pi
_t\left( q\right) +\left( d\pi _t\left( q\right) \right) ^{2} \\
&\equiv &\left( d\pi _t\left( q\right) \right) ^{2} \\
&=&-k^{2}\mathscr{V}_t^{2}\left( \beta \left( t\right) +\beta \left( t\right) ^{\ast
}\right) ^{2}dt.
\end{eqnarray*}
The first two terms $d\pi _t\left( q^{2}\right) +2\pi _t\left( q\right)
d\pi _t\left( q\right) $ cancel exactly, leaving an ODE for $\mathscr{V}_t$. 

We therefore obtain the following equation for the estimated position
observable:
\begin{eqnarray}
d\pi _t\left( q\right) =-k\mathscr{V}_t\left( \beta \left( t\right) +\beta \left(
t\right) ^{\ast }\right) \,dI\left( t\right) ,
\label{eq:filter_q}
\end{eqnarray}
where the conditional variance $\mathscr{V}_t$ satisfies the deterministic ODE
\begin{eqnarray}
\frac{d\mathscr{V}_t}{dt}=-k^{2}\mathscr{V}_t^{2}\left( \beta \left( t\right) +\beta \left(
t\right) ^{\ast }\right) ^{2}.
\label{eq:filter_var_q}
\end{eqnarray}
Note that $\mathscr{V}_t$ is decreasing so long as Re$\beta \left( t\right) \neq 0$,
and constant in any interval where Re$\beta \left( t\right) $ vanishes.

\bigskip 

We may further specify that the initial state is one where both canonical
coordinates $q$ and $p$ are jointly Gaussian. We may determine the filtered
estimate $\pi _t\left( p\right) $: first note that $e^{-i\theta
}pe^{i\theta }-p=2\hbar k$ and that
\begin{eqnarray*}
\mathcal{H}_t\left( p\right) =-k\left( \mathscr{C}_t-\frac{i}{2}\hbar
\right) \beta \left( t\right) -k\left( \mathscr{C}_t+\frac{i}{2}\hbar \right)
\beta \left( t\right) ^{\ast }
\end{eqnarray*}
where we introduce the symmetrized conditional covariance of $q$ and $p$ as
\begin{eqnarray}
\mathscr{C}_t \triangleq \frac{1}{2}\pi _t\left( qp+pq\right) -\pi _t\left( q\right)
\pi _t\left( p\right) .  
\label{eq:C_bar}
\end{eqnarray}
Therefore
\begin{eqnarray}
d\pi _t\left( p\right)& =&\hbar k|\beta \left( t\right) |^{2}dt \nonumber \\
&-&2k\mathscr{C}_t%
\text{Re}\beta \left( t\right) \,dI(t)-k\hbar \text{Im}\beta \left( t\right)
\,dI\left( t\right) .   \label{eq:pi_p}
\end{eqnarray}
Unlike the case of $\pi _t\left( q\right) $, we find a drift term
associated with $\pi _t\left( p\right) $ given by $\hbar k|\beta \left(
t\right) |^{2}dt$ which is interpreted as the momentum imparted by the
coherent source over the time interval $t$ to $t+dt$. To compute $\pi
_t\left( qp\right) $ we start with the filter equation for $X=qp$ which
reads as
\begin{eqnarray*}
d\pi _t\left( qp\right) =\hbar k\pi _t\left( q\right) |\beta (t)|^{2}dt+%
\mathcal{H}_t\left( qp\right) dI\left( t\right) 
\end{eqnarray*}
with
\begin{eqnarray*}
\mathcal{H}_t\left( qp\right)  &=&-k\left[ \pi _t\left( qpq\right) -\pi
_t\left( qp\right) \pi _t\left( q\right) \right] \beta \left( t\right) 
\\
&&-k\left[ \pi _t\left( q^{2}p\right) -\pi _t\left( qp\right) \pi
_t\left( q\right) \right] \beta \left( t\right) ^{\ast }
\end{eqnarray*}
and once again we may use (\ref{eq:3rd moments}) to break down the third
order moments. In fact, we obtain
\begin{eqnarray*}
\mathcal{H}_t\left( qp\right)  &=&-k\left[ \pi _t\left( q\right) \left( 
\mathscr{C}_t-\frac{i}{2}\hbar \right) -\pi _t\left( p\right) \mathscr{V}_t\right]
\beta \left( t\right)  \\
&&-k\left[ \pi _t\left( q\right) \left( \mathscr{C}_t+\frac{i}{2}\hbar
\right) -\pi _t\left( p\right) \mathscr{V}_t\right] \beta \left( t\right) ^{\ast }.
\end{eqnarray*}
From this we see that
\begin{eqnarray*}
&&d\left[ \pi _t\left( qp\right) -\pi _t\left( q\right) \pi _t\left(
p\right) \right]  \\
&=&d\pi _t\left( qp\right) -d\pi _t\left( q\right) \pi
_t\left( p\right)  \\
&&-\pi _t\left( q\right) d\pi _t\left( p\right) -d\pi _t\left(
q\right) d\pi _t\left( p\right)  \\
&\equiv &-k^{2}\mathscr{V}_t\left( \beta \left( t\right) +\beta \left( t\right)
^{\ast }\right)  \\
&\times& \left( \left[ \pi _t\left( qp\right) -\pi _t\left( q\right) \pi
_t\left( p\right) \right] \left( \beta \left( t\right) +\beta \left(
t\right) ^{\ast }\right) -i\hbar \beta \left( t\right) \right) dt.
\end{eqnarray*}
Once again, the $dI\left( t\right) $ terms cancel and we are left with a
deterministic ODE. Symmetrizing yields the deterministic equation
\begin{eqnarray}
\frac{d\mathscr{C}_t}{dt}=-k\mathscr{V}_t\mathscr{C}_t\left( \beta \left( t\right)
+\beta \left( t\right) ^{\ast }\right) ^{2}+\hbar \text{Im}\beta \left(
t\right) .
\label{eq:filter_C}
\end{eqnarray}

A similar computation works for the conditional uncertainty in the momentum
\begin{eqnarray}
\mathscr{W}_t \triangleq \pi_t (p^2 ) - \pi_t (p) ^2,
\end{eqnarray}
and we obtain the ODE
\begin{eqnarray}
\frac{ d \mathscr{W}_t }{dt} &=& (2 \hbar k)^2 ( \mathrm{Re} \beta (t) )^2 \nonumber\\
&&-(4k \mathscr{C}_t )^2
-16 \hbar k^2 \mathscr{C}_t \mathrm{Im} \beta (t) .
\end{eqnarray}

Note that the covariances come from a quantum Gaussian state, and so we must
have the inequality
\begin{equation*}
\left[ 
\begin{array}{cc}
\mathscr{V}_{t} & \mathscr{C}_{t}+\frac{i\hbar }{2} \\ 
\mathscr{C}_{t}-\frac{i\hbar }{2} & \mathscr{W}_{t}
\end{array}
\right] \geq 0
\end{equation*}
to be consistent with the Heisenberg uncertainty relations, see
for instance Section 3.3.3 in \cite{Eisert}.

\bigskip

\section{Conclusions}
We have derived the form of the filter (\ref{eq:interf_filter}) for the problem of 
estimating the quantum state of the a phase observable in an interferometer
based on detection of the output fields. As the photon fields do not interact
directly with the interferometer other than by scattering in the arms and
being split and recombined by the beam-splitters, we needed to place one of the
inputs at least in a non-trivial coherent state. This however lead to a practical
estimation problem. 

For the homodyne situation, we were able to work out the quantum Kalman-Bucy
filter. Here the conditional variance $\mathscr{V}_t$ evolves deterministically (\ref{eq:filter_var_q}).
If we make the modelling assumption that $\beta \left( t\right)
=\beta $ (constant) over the time interval of interest, then we obtain the explicit solution for $\mathscr{V}_t$ as 
\begin{eqnarray*}
\mathscr{V}_t=\frac{1}{V_{0}^{-1}+k^{2}\left( \beta +\beta ^{\ast }\right) ^{2}t}
\end{eqnarray*}
where $V_{0}$ is the variance of $q$ in the initial state $\rho _{0}$
assigned to the interferometer, i.e., $V_{0}=\mathrm{tr}\left\{ \rho
_{0}q^{2}\right\} -\left( \text{tr}\left\{ \rho _{0}q\right\} \right) ^{2}$.

The principal qualitative observation from this is, of course, that clearly $%
\lim_{t\rightarrow \infty }\mathscr{V}_t=0$. In other words, the conditional
variance is converging to zero as we acquire more information through the
quadrature measurement. What should happen in the long time asymptotic limit
is that, for any interval $A$, the probability of the observed position $q$
settling down to a value in $A$ will be given by tr$\left\{ \rho
_{0}P_{A}\right\} $ where $P_{A}$ is the projection operator 
\begin{eqnarray*}
\left( P_{A}\xi \right) \left( x\right) =\left\{ 
\begin{array}{cc}
\xi \left( x\right) , & x\in A; \\ 
0, & x\notin A.
\end{array}
\right.
\end{eqnarray*}
If the initial state $\rho _{0}$ was pure, corresponding to a wavefunction $%
\psi _{0}$, then the limit probability should be $\int_{A}|\psi _{0}\left(
x\right) |^{2}dx$. As far as we are aware, a rigorous proof of this
assertion is lacking, however it is well indicated for finite-dimensional
systems with discrete eigenvalues, see for instance \cite{SvHM04} and \cite{vHSM05}.

\end{document}